\documentclass[]{aa}
\usepackage{epsfig}
\begin{document}

\newcommand{\nc}{\newcommand}
\nc{\ALi}{$A_{\rm Li}$\,}
\nc{\Porb}{$P_{\rm orb}$\,}

  \title{ The tidal effects on the lithium abundance of binary
   systems with giant component\thanks{Based on observations 
  collected at ESO, La Silla.}  
                            }

  \author{  J. M. Costa, \inst{1}
            L. da Silva, \inst{2}
            J. D. do Nascimento Jr, \inst{1}
      \and  J. R. De Medeiros \inst{1}
  }
  \institute{
    Departamento de F\'{\i}sica, 
    Universidade Federal do Rio
    Grande do Norte, 59072-970 
    Natal,  R.N., Brazil
        \and
          Observat\'orio Nacional, 
          Rio de Janeiro, RJ, Brazil }

  \offprints{J. M. Costa}
  \mail{damata@dfte.ufrn.br}

  \date{Received / Accepted}


\abstract{ We analise the behavior of lithium abundance as a function of effective 
temperature, projected rotational velocity, orbital 
period and eccentricity for  a sample of 68 binary systems with giant component
and orbital period ranging from about 10 to 6400 days. For these binary systems the 
Li abundances show a gradual decrease with temperature, paralleling the well 
established result for single giants. 
We have also observed a dependence of lithium content on rotation. Binary systems 
with moderate to high rotation present also moderate to high Li content. This study 
shows also that synchronized binary systems with giant component seems to retain 
more of their original lithium than the unsynchronized systems. For orbital periods lower 
than 100 to 250 days, typically the period of synchronization for this kind of binary 
systems, lithium depleted stars seems to be unusual. The suggestion is made that there is 
an 'inhibited zone' in which synchronized binary systems with giant component
having lithium abundance lower than a threshold level should be unusual.
\keywords{ stars:   binaries --
           stars:     abundances --
           stars:     evolution  --
           stars:     interiors  --
           stars:     late-type
            }}

%

  \authorrunning{Costa, da Silva,  do Nascimento \& De Medeiros}
   \titlerunning{The tidal effects on the lithium abundance of binary systems}
\maketitle

\section{Introduction}

A correlation between the lithium content and the amount of angular 
momentum lost by late-type stars is predicted by different authors
(e.g: Pinsonneaut  et al.  1989, 1990; Zanh 1992, 1994). Such a prediction 
is indeed confirmed by observations. For unevolved late-type stars in young 
open clusters such as $\alpha-$Persei
and in Pleiades  there is a clear dependence of lithium abundance on
rotational velocity, in the  sense that the 
fastest rotators are generally stars with enhanced lithium content (Garcia 
L\'opez  et al.  1994; 
Randich  et al. 1998). For subgiant stars the same behavior was found by do
Nascimento   et al.  (2000), 
Randich  et al.  (1999) and De Medeiros  et al.  (1997), whereas
Barrado  et al.   (1998) confirmed 
such a correlation for chromospheric active binary systems. {\bf More recently 
De Medeiros  et al. (2000) have found the same behavior for a 
dependence of lithium content 
on rotation in giant stars, showing that single giants with  high
lithium content present also  high rotation rate}. 
Such a link between lithium 
depletion and angular momentum loss is also predicted for binary systems with 
late-type components. In close enough binary systems viscous dissipation of 
time-dependent 
tidal 
effects should produce a synchronization between rotation and stellar orbital 
motion as 
well as the circularization of the orbit of the system (e.g.: Zahn 1977). Thus, 
the angular momentum which is lost via stellar winds is drawn from the orbital 
motion, but 
with the result that the stars are spinning up, differently of single stars 
which are 
spinning down. In this context, for a given spectral type, a binary component 
will have a 
rotation enhanced in relation to its single counterparts and to those components 
of 
binary systems 
with orbital period larger than the critical period for synchronization. Hence, 
if the lithium  depletion is related to the loss of angular momentum, the 
surface
abundance of this element in 
synchronized binary systems should be less depleted than that of its single 
counterparts and that of nonsynchronized binary systems. 
In fact, Zahn (1994) has shown that late-type 
binary systems of 
short enough orbital period retain more of their original Li than their single 
counterparts. 
This author has found that such a period is typically below 8 days for 
solar-type stars of 
population I and below 6 days for halo stars. The same trend was found by
Spite   et al.  (1994) 
for old disk and halo stars. This inhibition of lithium depletion  was also 
observed in subgiant stars 
of population I (De Medeiros  et al.  1997, Randich  et al.  1999), 
whereas Barrado  et al. (1998) have found a similar result for giant components 
of chromospheric 
active binary systems. 
In the present work we {\bf analyse the effects of binarity on the lithium abundances 
of binary systems with evolved component of luminosity class III, 
typically F, G and K type giants, on the basis of high precision spectroscopic 
observations.}

\begin{figure}[t]
\vspace{.2in}
\centerline{\psfig{figure=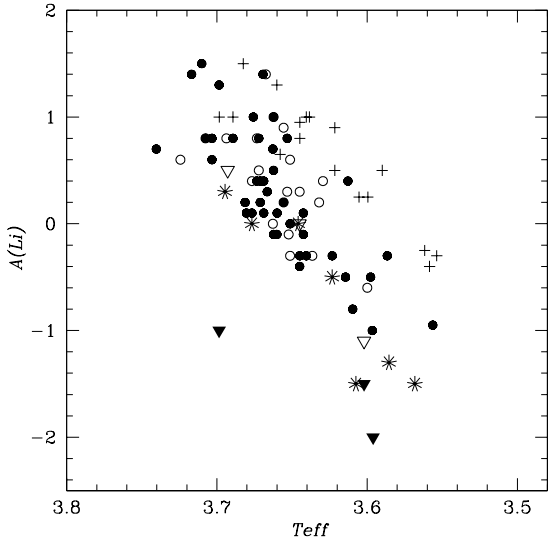,width=3truein,height=3truein}
\hskip 0.1in}

\caption[]{ \ALi as a function  of effective temperature.
Binary systems with  eccentricity lower than about 0.10 are identified by  
open circles, whereas filled  circles correspond to  systems with an eccentricity
 higher than 0.10. Filled inverted triangles are  upper  limits in the   
lithium abundance determination for systems with  eccentricity higher than 
about 0.10 and open   inverted   triangles are  upper limits
in the  lithium abundance determination for  systems with an eccentricity lower
 than 0.10.  Stars from NGC~7789 and M67 (Pilachowski  et al. 1988)
  are identified by cross and asterisks, respectively.}
\label{letterf2.ps}
\end{figure}

\begin{figure}[t]
\vspace{.2in}
\centerline{\psfig{figure=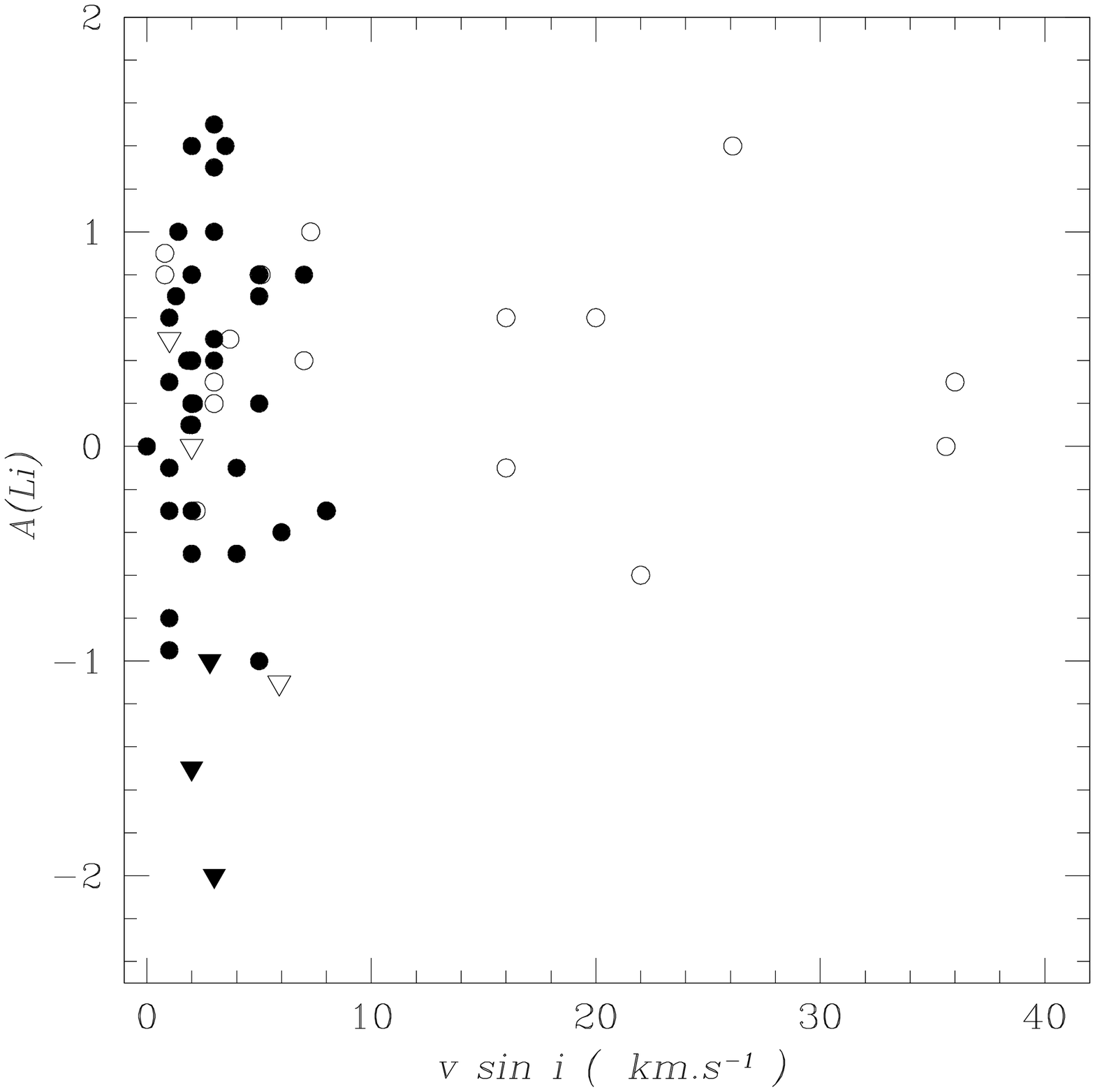,width=3truein,height=3truein}
\hskip 0.1in}

\caption[]{
\ALi as a function of rotational velocity.
Open symbols represent binary systems with
circularized orbit or nearly circularized orbit, namely those systems with
eccentricity lower than about 0.10, whereas the filled
symbols stand for the systems with an eccentricity higher than 0.10.
Inverted triangles are for upper limits in the  lithium abundance
determination.}
\label{letterf4.ps}
\end{figure}

\begin{figure}
\vspace{.2in}
\centerline{\psfig{figure=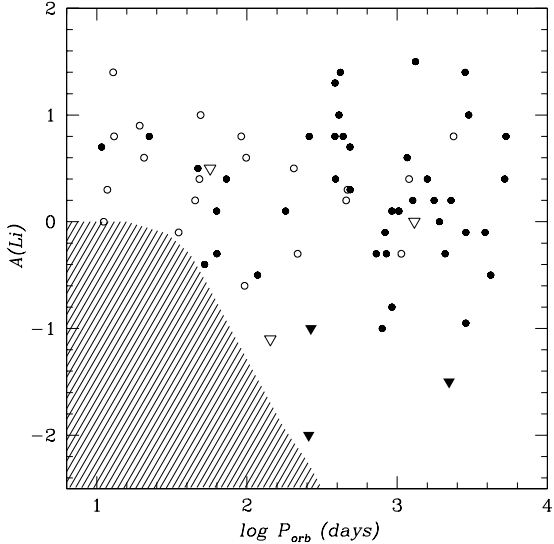,width=3truein,height=3truein}
\hskip 0.1in}

\caption[]{
\ALi as a function of the orbital period and eccentricity. 
Open symbols represent binary systems with
circularized orbit or nearly circularized orbit, namely those systems with
eccentricity lower than about 0.10, whereas the filled
symbols stand for the systems with an eccentricity higher than 0.10.
Inverted triangles are for upper limits in the  lithium abundance and
the shaded region defines a suggested 'inhibited zone', as explained in the
text.}
\label{letterf1.ps}
\end{figure}

\section{Working Sample and Observations}

The full sample analyzed in this study is composed by 68 binary systems with 
evolved component of luminosity class III, along the spectral region F, G and K, 
with 
orbital period ranging from about 10 to about 6400 days. The 
stars were selected because of their bright apparent magnitude, orbital 
parameters 
available in the 
literature and measured CORAVEL rotational velocity. We emphasize, however, that 
the sample 
is not complete in any particular aspect and we should not have inserted any 
bias with 
respect to some physical parameter. 
The Li I observations were obtained at ESO, La Silla, in two observing runs, 
January 1997 
and July 1998. The observations were carried out at the ESO CAT telescope 
equipped with the Coude Echelle Spectrometer. The covered spectral 
range was 56{\AA} 
wide, the resolving power was 95,000, linear dispersion 1.83{\AA}/mm
 and  signal-to-noise ratio 
between 80 and 200. Data reduction was performed by using the echelle context of 
the MIDAS package.  The \ALi were determined using a LTE code derived  
from one kindly made available for us by Monique Spite, from the Paris-Meudon 
Observatory, 
following the same 
procedure described in de la Reza and da Silva (1995), by fitting  the synthetic 
spectra to the observed ones. In the computation of the synthetic spectra,
all the known atomic lines in the range  $\lambda 6702$\,--\,$\lambda 6712$ were 
considered. The hyperfine structure of the lithium resonance line was taken into 
account using the data 
quoted by Duncan (1991). The stellar atmospheric models used are from Gustafsson 
and collaborators 
(Bell  et al.  1976). For all the stars we used the log g value given by 
the 
spectral type,
microturbulence velocity of \mbox{2 km s$^{-1}$} and  solar
abundances. Special care should be taken for the estimation of the effective 
temperature, since the 
derivation of  \ALi is very sensitive to this parameter. For stars with 
the Stromgren ubvy photometry we have used the calibration from Moon (1985),
where   the effective temperature and surface gravity are given as a function
of the ubvy   index. For 
those stars with no available Stromgren photometric index we have used the 
calibrations from 
Flower (1977) where the effective temperature is given as a function of the 
(B-V) index and 
that from Pasquini  et al.  (1990) which gives the effective 
temperature as a 
function of 
the (V-R) index. {\bf The derived lithium abundances as well as main stellar parameters 
and rotational velocities from CORAVEL, are presented in Table 1. Before discussing the 
results of the present study let us analyse how accurate are our lithium abundances 
and rotational velocities. First of all, we have compared the effective temperature 
values computed in the present work with those available in the literature for common stars, 
obtained by different authors. From 16 stars of  Barrados et al. (1998), 6 stars of  
Brown  et al. (1989) and 8 stars of Randich  et al. (1994), in 
common with our sample, least  
square fits of the temperature values give, respectively, coefficients of correlation of 
about 0.90, 0.96 and 0.94 as well as  r.m.s of the differences of about 140~K, 104~K  
and 60~K, 
respectively, indicating that our derived temperatures are in very reasonable agreement 
with others. Finally, we have compared Li abundances between the present study and the 
results of previous investigations  cited above. 
For the 16 stars in common with Barrados et al. (1998) a least square fit  give a 
correlation coefficient of about 0.80. A comparison with the Li abundances from Brown  et al.
 (1989) and Randich  et al. (1994) gives a coefficient of correlation 
better than 0.80.}

\section{Results and Discussion} 

{\bf In Fig. \ref{letterf2.ps} we show the behavior of the Li abundance for the binary systems with evolved 
component as a function of the effective temperature. The gradual decrease 
of lithium content with effective temperature, well established for single giants 
(e.g.: Brown  et al. 1989), is clearly observed for the 
binary systems. In addition, we are 
displaying in the Fig. \ref{letterf2.ps} the Li abundances for the open clusters NGC 7789 and M 67 from 
Pilachowski  et al. (1988), with ages of about 1.7 Gyrs and 5.0 Gyrs respectively. 
A comparison of the Li abundances of the present study with those for the cited open clusters 
shows, at a given effective temperature, a clear decrease of Li content with age for the 
binary systems with evolved component. Fig. \ref{letterf4.ps} presents the behavior of the Li abundances versus 
the projected rotational velocity $v\sin~i$. Two interesting features emerge from this 
figure, paralleling the same link between rotation and lithium content for single giant stars
(De Medeiros  et al. 2000). First, one observes a large spread in the  
distribution of the values of Li abundance at 
low values of the rotational velocity. For binary systems with $v\sin~i$ lower than about 
5 km~s$^{-1}$, there is a spread of Li abundances from about -2 to about 1.5. Such a spread
decreases  with increasing rotation. Second, binary systems with moderate to high $v\sin~i$ values, 
present also moderate to high values of Li abundance, this result pointing 
for a correlation between rotation and Li content.
Fig. \ref{letterf1.ps} shows lithium abundance as a function of orbital period for 
binary systems with giant component. It is clear that for binary systems with 
an orbital period lower than about 
the expected period of synchronization, typically a period between 100 and 250 
days, there is a strong inhibition in the lithium depletion.   
As indicated by the shaded area represented in the Fig. \ref{letterf1.ps}, 
such an 'inhibited zone' is located below  \ALi values around 
0.0 and to the left of the expected values of the orbital period of 
synchronization.  For the unsynchronized systems, namely those with an orbital period greater than 
about 100 to 250 days, one observes a large spread in Li abundances, with  \ALi 
ranging from about -2 to about 1.5, in contrast to synchronized systems 
which present a narrow spread in  \ALi from about 0.0 to at least 1.5. 
Our finding is reinforced by the results of Barrado  et al.  (1998) for 
chromospherically active binary systems with giant component. By analyzing 
the lithium behavior of 66 such systems, most of them presenting synchronization 
features, with an orbital period lower than about 100 
days, those authors have found no stars showing lithium abundance significantly 
below such a critical value of  \ALi  around 0.0, as represented in their
 Fig. 6b. Further, 
Barrado  et al.  (1998) list a few chromospherically active 
binary systems with giant component and orbital period shorter than 100 days 
showing \ALi around the cosmic value 3.0.    
Because the synchronization between rotational and orbital motion results from 
tidal effects, we claim that the 'inhibited zone' is the result of the inhibition of 
the depletion of lithium due to such tidal effects. Hence, whereas synchronized 
binary systems have a tendency to retain more of their original lithium, the 
unsynchronized systems have normal lithium depletion. This explains the 
'inhibited zone', which indicates the absence of synchronized binary systems 
with giant component having a lithium abundance lower than a threshold level in 
the \ALi - \Porb plane.

\begin{table*}[t]
\small
\caption{Lithium abundances, rotational velocities and orbital parameters
for  binary systems with evolved  component
}
\label{basej1}
\begin{flushleft}
\begin{tabular}{rrcrrrrc}
\noalign{\smallskip}
\hline\\[-2mm]

\multicolumn{1}{c}{HD}&
\multicolumn{1}{c}{ST}&
\multicolumn{1}{c}{B$-$V}&
\multicolumn{1}{c}{{$v\sin~i$}}&
\multicolumn{1}{c}{\ALi}&
\multicolumn{1}{c}{\Porb}&
\multicolumn{1}{c}{$e$}&
\multicolumn{1}{c}{Ref}
\\[2mm]
\hline\\[-3mm]
\\
28     & K1III  & 1.04 &   3.0   &  0.4   &  72.93  &  0.27  &   a       \\
352    & K2III  & 1.38 &   22.0  &  $-$0.6  &  96.439  & 0.04  & a \\
1833   & K1III  & 1.13 &   16.3  &  $-$0.1  &  35.100  & 0.04  & b \\
2261   & K0III  & 1.09 &   1.0   &  $-$0.1  &  3848.83 & 0.34  & a \\
7672   & G5IIIe & 0.90 &   1.0   &  0.5   &  56.8147 & 0.04  &   a \\
12923  & K0III  & 0.90 &   2.0   &  0.8   &  5302   & 0.432  &  e \\
19754  & G8III-IV & 1.12&  7.0   &  0.4   &  48.263 & 0.1    &  c \\
22905  & G8III  & 0.88 &   0.8   &  0.8   &  91.629 & 0.0    &  a \\
23817  & K2III  & 1.13 &   0.0   &  0.0   &  1911.5 & 0.21   &  a \\
34802  & K1IIIp & 1.09 &   0.8   &  0.9   &  19.310 & 0.0    &  b \\
37297  & G8-K0III & 0.83&  2.0   &  0.1   &  180.87 & 0.51   &  a \\
38099  & K4III  & 1.47  &  5.9   &  $<~-$1.1  &  143.03 & 0.06   & a \\
43821  & G5III  & 0.87  &  3.0   &  1.5   &  1325   & 0.44   &  a \\
46407  & K0III  & 1.11  &  3.0   &  0.2   &  457.7  & 0.0    &  a \\
49293  & K0IIIa & 1.11  &  2.0   &  0.2   &  1760.9 & 0.40   &  a \\
50310  & K1III  & 1.20  &  2.2   &  $-$0.3  &  1066.0 & 0.09  &  a \\
58972  & K3III  & 1.22  &  1.8   &  0.4   &  389.0  & 0.31   & a \\
59717  & K5III  & 1.52  &  3.0   &  $<~-$2.0  &  257.8  & 0.17   &  a \\
61245  & K2III  & 1.16  &  36.0  &  0.3   &  11.761 & 0.01   &  b \\
79910  & K2III  & 1.17  &  2.0   &  0.1   &  922    & 0.29   &  a \\
81410  & K1III  & 1.02  &  26.1  &  1.4   &  12.8683 & 0.0   &  a \\
82674  & K0III  & 1.17  &  1.0   &  $-$0.1  &  830.4  & 0.15  & a \\
83240  & K1III  & 1.05  &  2.0   &  1.4   &  2834   & 0.32  & a \\
83442  & K2IIIp & 1.16  &  6.0   &  $-$0.4  &  52.270 & 0.13  & b \\
88284  & K0III  & 1.01  &  1.9   &  0.4   &  1585.8 & 0.14   & a \\
92214  & G7.5III& 0.92  &  2.0   &  0.4   &  1200   & 0.1    &  a \\
94363  & K0III+ & 0.90  &  1.0   &  0.6   &  1166   & 0.38    & a \\
102928 & K0III  & 1.06  &  1.0   &  0.3   &  486.7  & 0.31     & a \\
112048 & K0III  & 1.09  &  1.9   &  0.1   &  1027   & 0.32    & a \\
112985 & K2III  & 1.18  &  1.0   &  $-$0.3  &  847    & 0.4    &  a \\
119834 & G9III  & 0.96  &  2.0   &  0.8   &  437.00 & 0.13     & a \\
120901 & K0III  & 1.08  &  5.0   &  0.2   &  2283.0586 & 0.49  &  a \\
133461 & K2III  & 1.16  &  2.0   &  $-$0.3  &  725.5  & 0.29    & a \\
136905 & K1III  & 1.03  &  35.6  &  0.0   &  11.1345 & 0.00  & a \\
139137 & G8III  & 0.72  &  2.0   &  0.8   &  259.81 & 0.378  & f \\
145206 & K4III  & 1.45  &  2.0   &  $-$0.3  &  2084.8 & 0.55  & a \\
147508 & K2III  & 1.33  &  1.0   &  $<~-$0.8  &  922.8  & 0.37   & a \\
156731 & K3III  & 1.45  &  5.0   &  $-$1.0  &  794.5  & 0.69  & a \\
158837 & G8III  & 0.84  &  3.5   &  1.4   &  418.242 & 0.20   & a \\
162391 & G8III  & 1.13  &  8.0   &  $-$0.3  &  217.440 & 0.0    & d \\
162596 & K0III  & 1.12  &  3.0   &  0.3   &  467.2   & 0.0    &  a \\
165141 & G8-K0III & 1.01&    2.0 &    0.4 &    5200. & ...   & c \\
168339 & K4III  & 1.48  &  2.0   &  $<~-$1.5  &  2214  &  0.26  &   a \\
169156 & G9IIIb & 0.94  &  5.0   &  0.8   &  2373.7911 &0.10  &   a \\
169689 & G8III-IV & 0.92 & 7.0   &  0.8   &  385    & 0.31  &   a \\
169985 & G0III  & 0.50   & 3.0   &  1.3   &  386.0  & 0.47  &  a \\
172831 & K0-1III & 1.00  &  1.3  &   0.7  &   485.3 &  0.21 &   a \\
175515 & K0III  & 1.04   & 1.4   &  1.0   &  2994   & 0.24  &   a \\
176411 & K1III  & 1.08   & 2.1   &  0.2   &  1270.6 & 0.27  &   a \\
178717 & K4III  & 1.88   & 1.0   &  $-$0.95 &  2866   & 0.434 &   g \\
179950 & F2:+F1III& 0.55 & 5.0   &  0.7   &  10.7786 & 0.47   & a \\
181391 & G8III-IV & 0.92 & 2.8   &  $<~-$1.0  &  266.544 & 0.83 &  a \\
181809 & K2III & 1.17   & 5.1    &  0.8   &  13.048 &  0.05 &    c \\
182776 & K2/K3III & 1.17 & 2.0   &  0.2   &  45.180 & 0.02  &   c \\
185510 & KOIII-IV & 0.84 & 16.0  &  0.6   &  20.660 & 0.10  &   b \\
\\
\hline
\end{tabular}
\end{flushleft}
\end{table*}

\begin{table*}[t]
{\bf Table 1}.(continued)
\small
\label{basej1}
\begin{flushleft}
\begin{tabular}{rrcrrrrc}
\noalign{\smallskip}
\hline\\[-2mm]

\multicolumn{1}{c}{HD}&
\multicolumn{1}{c}{ST}&
\multicolumn{1}{c}{B$-$V}&
\multicolumn{1}{c}{{$v\sin~i$}}&
\multicolumn{1}{c}{\ALi}&
\multicolumn{1}{c}{\Porb}&
\multicolumn{1}{c}{$e$}&
\multicolumn{1}{c}{Ref}
\\[2mm]
\hline\\[-3mm]
\\
188981 & K1III  & 1.05   & 2.0   &  0.1   &  62.877 & 0.34  &   a \\
194184 & K3III  & 1.36   & 4.0   &  $-$0.5  &  117.776 & 0.24 &  a \\
196574 & G8III  & 0.95   & 3.7   &  0.5   &  205.2  & 0.0   &   a \\
202134 & K1IIIp & 1.12   & 8.0   &  $-$0.3  &  63.09  & 0.52  &  c \\
204128 & K1III  & 1.12   & 5.0   &  0.8     &   22.349 &  0.12 &  c \\
202447 & G0III$+$ & 0.53   & 20.0  &  0.6   &  98.81  & 0.04  &   a \\
205249 & K1IIIp & 1.08   & 7.3   &  1.0   &  49.137 & 0.08  &   c \\
205478 & K0III  & 1.00   & 2.0   &  0.1   &  1020   & 0.4   &   a \\
211416 & K3III  & 1.39   & 2.0   &  $-$0.5  &  4197.7 & 0.39  &   a \\
213428 & K0III  & 1.08   & 4.0   &  $-$0.1  &  2866   & 0.5   &   a \\
217188 & K0III  & 1.08   & 3.0   &  0.5   &  47.121 & 0.50  &  c \\
218670 & K1III  & 1.02   & 3.0   &  1.0   &  409.614 & 0.66 &    a \\
223617 & G9III  & 1.16   & 2.0   & $<$ 0.0   &  1301  &  0.098 &  g \\

\\
\hline
\end{tabular}
\end{flushleft}
a - Batten et al. (1989); 
b - Strassmeier  et al. (1988);
c - Strassmeier  et al. (1993);
d - Mermilliod  et al. (1989);
e - Griffin (1989);
f - Griffin (1990);
g -  McClure et al. (1990)

\end{table*}

\section{Conclusions} 

Lithium abundances are presented for 68 binary systems with giant component. By combining 
these data with effective temperature, projected 
rotational velocity $v\sin~i$ and orbital parameters, we show some very interesting 
trends on the behavior of Li content in this class of binary systems. The distribution 
of Li abundance as a function of effective temperature follows the same behavior observed 
for single giants, namely a gradual decrease of Li content with effective temperature. At low values 
of rotational velocity, typically for $v\sin~i$ lower than about 5 km~s$^{-1}$, there is 
a large spread in the values of Li abundance of at least 3.5 magnitudes. Such a 
spread decreases for increasing rotation, following the same trend observed for 
their single counterparts. In spite of this large spread in Li abundances for 
the slow rotators, we observe that the moderate to high rotators present a 
tendency for moderate to high 
lithium contents, this result pointing for a correlation between rotation and Li content.  
Finally, the analysis of the Li abundance as a function of orbital period seems to show 
the effect of tidal interaction on the dilution of lithium. Binary systems with orbital 
period lower than about 100 days, typically those systems showing synchronization between 
rotational and orbital motions, present Li abundance enhanced in relation to the systems 
with orbital period larger than 100 days. In fact, in the  \ALi - \Porb plane it seems 
to exist an 'inhibited zone' where binary systems with giant component
showing synchronization between rotational and orbital motions and abundances 
lower than a threshold value around 0.0 appear to be unusual.

\begin{acknowledgements}
This work has been supported by continuous grants from the CNPq Brazilian 
Agency. L. da S. 
thanks the CNPq for financial support through grant 200580/97-0. 
J.D.N.Jr. acknowledges the CNPq grant 300925/99-9.      
We would like to thank the referee for his 
useful comments  and suggestions   on the
manuscript.
\end{acknowledgements}


\begin{thebibliography}{}

\bibitem[]{Barrado98}
Barrado y Navascues, D., de Castro, E., Fernandez-Figueroa, M. J., 
Cornide, M., Garcia L\'opez, R. J. 1998, A\&A 337, 739


\bibitem[]{Batten89}
Batten, A. H., Fletcher, J. M., MacCarthy, D. G. 
1989, Publ. Dom. Astrophys. Obs. vol. 17, 1


\bibitem[]{Bell76}
Bell, R. A., Eriksson, K., Gustafsson, B., Nordlund, A.  1976,
A\&AS   23, 37
 
\bibitem[]{Brown89}
Brown, J. A., Sneden, C., Lambert, D. L., Dutchover, E. Jr. 1989, ApJS 71, 293

\bibitem[]{DeMedeiros97}
De Medeiros, J. R., do Nascimento, J. D. Jr., Mayor, M. 1997, A\&A 317, 701

\bibitem[]{DeMedeiros2000}
De Medeiros, J. R.,  do Nascimento, J. D. Jr.,  Sankarankutty, S.,
Costa, J. M., Maia, M. R. G.  2000, A\&A 363, 239


\bibitem[]{LaReza95}
  de la Reza, R., da Silva, L.  1995, ApJ 439, 917

\bibitem[]{doNascimento2000}
  do Nascimento, J. D. Jr., Charbonnel, C., L\`ebre, A., de Laverny, P.,
  De Medeiros, J. R.  2000, A\&A 357, 931

\bibitem[]{Duncan}
  Duncan, D. K. 1991  ApJ 373, 250

\bibitem[]{Flower77}
 Flower, P. J. 1977 A\&A 54, 31

\bibitem[]{Garcia94}
Garcia Lopez, R. J., Rebolo, R., Martin, E. L. 1994, A\&A 282, 518


\bibitem[]{Griffin89}
Griffin, R. F.  1989, The Observatory 109, 142 

\bibitem[]{Griffin90} 
Griffin, R. F. 1990, JApA 11, 491


\bibitem[]{McClure90}
McClure, R. D., Woodsworth, A. W. 1990, ApJ 352, 709

\bibitem[]{Moon85}
Moon, T. 1985, Comm. University of London Observatory No. 78

\bibitem[]{Mermilliodetal89}
Mermilliod, J.-C., Mayor, M., Andersen, J.,
 Nordstrom, B., Lindgren, H., Duquennoy, A. 1989, A\&AS 79, 11

\bibitem[]{Pasquini1990}
 Pasquini, L., Brocato, E., Pallavicini, R. 1990, A\&A 234, 277


\bibitem[]{Pila1988}
 Pilachowski, C. A., Saha, A., Hobbs, L. M. 1988, PASP 100, 474
 

\bibitem[1989]{Pinsonneault89}
 Pinsonneault, M. H., Kawaler, S. D., Sofia, S., Demarque, P. 1989, ApJ 338, 424 

\bibitem[1990]{Pinsonneault90}
  Pinsonneault, M. H., Kawaler, S. D., Demarque, P. 1990, ApJS 74, 501 

\bibitem[1994]{Randich1994}
Randich, S., Giampapa M. S., Pallavicini R. 1994, A\&A 283, 893

\bibitem[]{Randich1998}
Randich, S., Martin, E. L., Garcia L\'opez, R. J., Pallavicini, R.
 1998, A\&A 333, 591 

\bibitem[]{Randich1999}
Randich, S. Gratton, R., Pallavicini, R., Pasquini, L., Carretta, E. 1999, A\&A 
348, 487

\bibitem[]{spite94}
  Spite, M., Pasquini, L., Spite, F. 1994, A\&A 290, 217

\bibitem[]{Strassmeier88}
Strassmeier, K. G., Hall, D. S., Zeilik, M., Nelson, E.,
 Eker, Z., Fekel, F. C.  1988, A\&AS 72, 291


\bibitem[]{Strassmeier93}
 Strassmeier, K.G., Hall, D.S., Fekel, F.C., Scheck, M. 1993, A\&AS  100, 173



\bibitem[]{zahn77}
  Zahn, J. P. 1977, A\&A 57, 383


\bibitem[]{zahn94}
  Zahn, J. P. 1994, A\&A 288, 829

\end{thebibliography}
\end{document}